\documentclass[journal,draftcls,onecolumn,12pt,twoside]{IEEEtranTCOM}%\documentclass[final,,twoside]{IEEEtranTCOM}

\normalsize

\ifCLASSINFOpdf

\else

\fi

\normalsize
\ifCLASSINFOpdf
\else
\fi

\usepackage{amsthm}
\usepackage{amsmath}
\usepackage{amssymb}
\usepackage{graphicx}
\usepackage{esint}
\usepackage{psfrag}
\usepackage{amsfonts}
\usepackage{cite}
\usepackage{balance}
\usepackage{subcaption}
\usepackage{epstopdf}
\usepackage{color}
\usepackage{afterpage}
\usepackage{hyperref}

\makeatletter

\theoremstyle{plain}

\makeatother

\providecommand{\theoremname}{Proposition}

\renewenvironment{IEEEbiography}[1]
  {\IEEEbiographynophoto{#1}}
  {\endIEEEbiographynophoto}

\begin{document}
%\markboth{\textit{IEEE Signal Processing Letters}}{Boulogeorgos, et al.:  Optimal Power Allocation for OFDMA Systems Under I/Q Imbalance}
%\markboth{\textit{IEEE Communications Magazine}}{Boulogeorgos, et al.:  Terahertz Technologies to Deliver Optical Network Quality of Experience in Wireless Systems Beyond~5G}
\title{Terahertz Technologies to Deliver Optical Network Quality of Experience in Wireless Systems Beyond~5G}

\author{Alexandros-Apostolos A. Boulogeorgos, 
Angeliki Alexiou,
%\\
Thomas Merkle,
Colja Schubert, 
Robert Elschner, 
Alexandros Katsiotis,
Panagiotis Stavrianos,
Dimitrios Kritharidis,
Panteleimon-Konstantinos Chartsias, 
Joonas Kokkoniemi,
Markku Juntti,
Janne Lehtom\"{a}ki,
Ant\'onio Teixeira, and
Francisco~Rodrigues

\thanks{A.-A. A. Boulogeorgos and A. Alexiou are with the department of Digital Systems, University of Piraeus, Piraeus 18534, Greece. (E-mails: al.boulogeorgos@ieee.org, alexiou@unipi.gr).}
\thanks{T. Merkle is with the Fraunhofer Institute of Applied Solid State Physics (IAF), Freiburg, 79108, Germany. (E-mail: thomas.merkle@iaf.fraunhofer.de).}
\thanks{C. Schubert and R. Elschner are with the Fraunhofer Institute for Telecommunications, Heinrich Hertz Institute (HHI), Berlin, 10587, Germany. (E-mail: \{colja.schubert, robert.elschner\}@hhi.fraunhofer.de).}
\thanks{A. Katsiotis, P. Stavrianos, D. Kritharidis and P.-K. Chartsias are with the Intracom Telecom, 19.7 Km Markopoulou, Ave., 190 02 Peania, Greece. (E-mails: \{alexkat, panstav, dkri, chkopa\}@intracom-telecom.com).}.
\thanks{ J. Kokkoniemi, M. Juntti, and J. Lehtom\"{a}ki are with the Centre for Wireless Communications (CWC), University of Oulu, 90014 Oulu, Finland. (E-mail:  \{joonas.kokkoniemi, markku.juntti, janne.lehtomaki\}@oulu.fi).}.
\thanks{A. Teixeira and F. Rodrigues are with the PICadvanced S.A., IEUA Campus Universit\'ario Santiago Ed.1 Sala 18 3810-193 Aveiro, and University of Aveiro (E-mails: \{teixeira, francisco\}@picadvanced.com).}
}
\maketitle	

\vspace{-2cm}
%\newpage
\begin{abstract}
%The concept of extending the fiber-optic systems quality of experience and reliability to wireless access and backhaul links in networks beyond the {fifth} generation (5G) era is expected to play a decisive role in supporting novel bandwidth-hungry applications. 
This article discusses the basic system architecture for  {terahertz (THz)} wireless links {with} bandwidths of more than $50$ GHz into optical networks. New design principles and breakthrough technologies are required {in order to demonstrate Tbps data-rates}  at near zero-latency using the proposed system concept. Specifically, we present {the concept} of designing the baseband signal processing for both the optical and wireless link and using an end-to-end (E2E) error correction approach for the combined link. We provide two possible electro-optical baseband interface architectures, {namely}  transparent optical-link and  digital-link {architectures}, which are currently under investigation. THz wireless link requirements are given as well as the main principles and research directions for the development of a new generation of {transceiver} frontends, which {will be} capable of operating at ultra-high spectral {efficiency} by employing higher-order modulation schemes. Moreover, {we discuss the need for developing a novel THz network information theory framework, which {will take} into account the channel characteristics and the nature of interference in the THz band.} Finally, we highlight the role of pencil-beamforming (PBF), which is required in order to overcome the propagation losses, as well as the physical layer and medium access control challenges.
\end{abstract}

%\newpage
%\begin{IEEEkeywords}
%Beyond 5G, Co-design principle, H-band, Terahertz communications. 
%\end{IEEEkeywords}

\section{Introduction}\label{S:Intro}
Over the last years, the proliferation of wireless devices and the increasing number of {bandwidth-consuming internet} services have significantly raised the demand for high data-rate transmission with very low latency. While the wireless world {moves} towards the fifth generation (5G) era, several technological advances, such as massive multiple-input multiple-output (MIMO) {systems}, full duplexing, and millimeter wave (mmW) communications, have been {presented} as promising enablers. However, there {is} a lack of efficiency and flexibility in handling the huge amount of quality of service (QoS) and experience (QoE) oriented~data.

{In view of the fact that the currently used frequency spectrum for 5G has limited capacity, THz wireless {became} an attractive complementing technology to the less flexible and {more expensive} optical-fiber connections {as well as} to the lower data-rate {systems}, such as visible light communications, microwave links, and WiFi~\cite{6005345,6892933}. }
As a consequence, THz {communications are expected to be used} for wireless access and  backhaul {networking; hence, they will influence} the main technology trends in wireless networks within the next ten years and beyond. 
The implementation of THz  networks will have to leverage breakthrough novel technological concepts. 
Examples are the joint-design of baseband digital signal processing (DSP) for the complete optical and wireless link, the development of broadband and highly spectral efficient radio frequency (RF) frontends operating at frequencies higher than $275$ GHz, and new standardized electrical-optical (E/O) interfaces. 
{Additionally, } to address the extremely large bandwidth and the propagation properties of the THz regime, improved channel modeling and the design of appropriate waveforms, {multiple access control (MAC)} schemes and antenna array configurations {are} required.

Motivated by the potential of THz technologies, this article focuses on illustrating the vision and approach of delivering optical network QoE in wireless systems, presenting the critical technology gaps, {and} the appropriate enablers. {Such technologies will bring ``fiber-optic speeds out of the fiber'' offering Tbps wireless connectivity. Also, it will exceed nowadays $100$ Gbit/s target by a factor of $10$ by {dramatically} increasing the spectrum efficiency.}
The key idea is the {co-design} of the baseband DSP for both the optical and wireless links in the combined optical-wireless network architecture.
 Setting out from this idea, {several} of the current design principles and architectures need to be redefined, mainly by replacing the notion of joint-optimization with that of `co-design', {e.g.,} optical and wireless, backhaul and access, channel models and waveforms, signals and coding, beam-patterns and MAC schemes. In this context, the identification of the critical technology gaps and invention, optimization and demonstration of the appropriate enablers, are expected to become the fundamental pillars, which will {catalyze} the road  beyond~5G. 

\section{Beyond 5G Scenarios, Applications and Requirements}\label{S:Beyond5G}

\begin{figure}
\centering\includegraphics[width=1.04\linewidth,trim=0 0 0 0,clip=false]{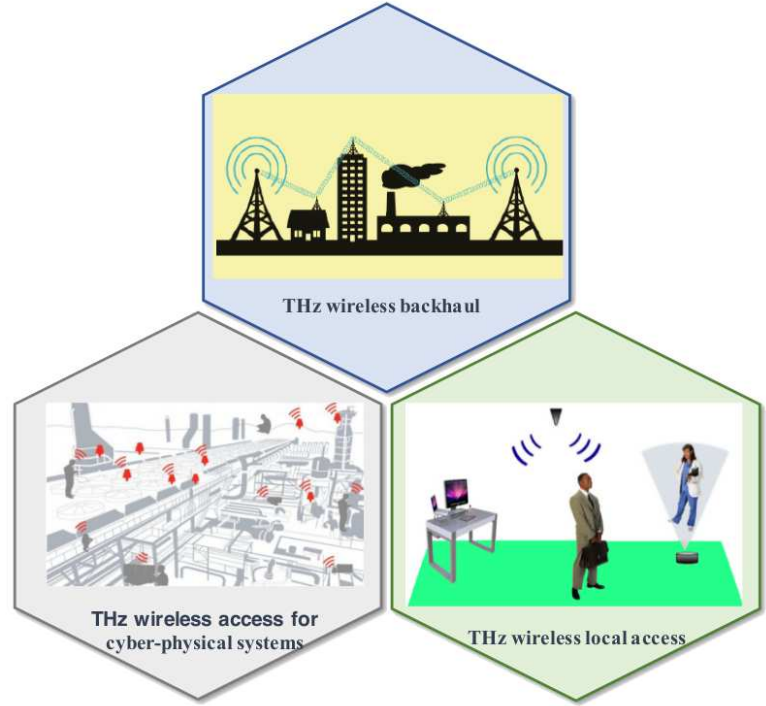}
%\vspace{-1.1cm}
\caption{Application scenarios.}
\label{fig:Scenarios}
\end{figure}

Networks beyond 5G are envisioned to provide unprecedented performance excellence, not only by {targeting Tbps data-rates,} but also by inherently supporting a large range of novel usage scenarios and applications that combine \textit{Tbps data-rates with agility, reliability, and zero response time}. Virtual presence, 3D printing, cyber physical systems for intelligent transport, and industry 4.0 are only a few examples of several highly challenging anticipated use cases. In order to enhance scalability, flexibility and efficient resources allocation, 5G  embrace several game changing design approaches, such as virtualization, softwarization, orchestration and commoditization of resources. However, these approaches do not tackle the fundamental performance limitations related to the available bandwidth, transmission and processing delay, cost and energy consumption, which still define the envelope of 5G capabilities. To break these barriers in networks beyond 5G, {we} need to bring little-explored resources and technologies to validation by directing research towards de-risking technological concepts, components, architectures and systems concepts from an early design stage on. {Hence}, innovative joint-investigation, assessment and design of theoretical models, aligned and supported by experimental parameter {evaluation/estimation} and validation are~required. {Next, we present the main application scenarios and their key requirements. These scenarios are illustrated in Fig.~\ref{fig:Scenarios}.}  

\subsubsection*{THz wireless backhaul} In the future, the users in rural or remote regions, which nowadays suffer from low-connectivity, should {enjoy} $10$ Gbps data-rates. Fiber-optical solutions are {often} time-consuming, since one {should} wait to capitalize scheduled road-reconstructions {and anyway} very costly. 
{Therefore,} the connectivity to rural areas {does  not often} progress at all. 
However, the access to high-speed internet is a crucial advantage in the global competition for industry sites and highly qualified human resources. 
If a cost-efficient and flexible {solution} cannot be guaranteed, the {digital divide} between rural areas and major cities will increase. THz {links}, as a wireless backhaul extension of the optical-fiber, is an important building block to guarantee high-speed internet access everywhere. Moreover, the increasing number of mobile and fixed users in {both the private, industrial and service sectors} will require hundreds of Gbps in the communication to/or between cell towers (\textit{backhaul}) or between cell towers and remote radio heads (\textit{fronthaul}). In such scenarios, apart from the high {targeted} data-rates ($\sim 1$ Tbps), the critical parameter is~\textit{range}, {which should be in the order of some~kilometers.}

\subsubsection*{THz wireless local access} THz communications will enable seamless {connectivity} between ultra-high-speed wired networks and personal wireless devices, achieving full transparency and rate convergence between {the two} links. This will facilitate the use of bandwidth-hungry applications across static and mobile users, mainly in \textit{indoor and local access scenarios}. Some specific applications are high-definition holographic video conferencing  or ultra-high-speed wireless data-distribution in data-centers. The critical parameter {here,} apart from the Tbps data-rates, is \textit{connection reliability} {with bit error rates in the order of $10^{-5}$}, in a low mobility~environment.

\subsubsection*{THz wireless access for cyber physical systems} Fully adopting digital networking in industry, commerce and public services, including traffic control and autonomous driving, remote health monitoring services, supply chain, security and safety procedures, automation of large production sites, place stringent requirements for Tbps-class access subject to fast response constraints. These \textit{cyber physical} scenarios describe the true colors of {the} commonly-known Tactile Internet {and}  challenge the capability of systems beyond 5G to offer \textit{almost zero latency} {($\sim 1$ ms)}, apart from the high data-rates {in} the order of~Tbps.

\section{THz Wireless-Optical Radio Systems} \label{S:THzBandCharacteristics}

It is intuitive that data-rate compatibility is required to extend the QoS and QoE of fiber-optic systems over a THz wireless link. 
{Recently,} coherent fiber-optical data transmission has pushed more and more into short distance application areas, previously occupied by non-coherent optical systems. The main drivers are the progress in optical transponder technologies and the associated decreasing component costs. The upcoming {commercially} available transponder chipset generation for coherent fiber transmission will support symbol rates up to $64$ Gbd and modulations of $64$-QAM {\cite{roberts2017}}. {Higher symbol rates and modulations up to 256-QAM or even 1024-QAM at $64$~Gbd have been experimentally demonstrated in recently published testbeds over distances of $400$ km and data-rates of $1.32$ Tbps~\cite{maher2017}}. 

The progress of wireless THz transceiver frontends has {provided} bandwidths of more than $50$ GHz at frequencies above $275$ GHz~{\cite{C:Testbed}}. The fiber-optical transponder {and} the wireless THz transceiver architectures resemble each other, since both use coherent transmission in combination with quadrature architectures~{\cite{roberts2017}}. Blending these two technologies seems to be a plausible strategy. While there has been work  on applying RF-over-Fiber (RoF) concepts to this problem, this article is rather focusing on the approach of aligning with optical transponder technologies from coherent fiber-optics, which is introduced as `photonic radio' to clearly distinguish from~RoF. {This approach offers the possibility to exploit the technological progress in modems for optical coherent transceivers and associated advances in DSP for the implementation of THz wireless~links.}

\subsection{System Overview}

\begin{figure}
\centering\includegraphics[width=1\linewidth,trim=0 0 0 0,clip=false]{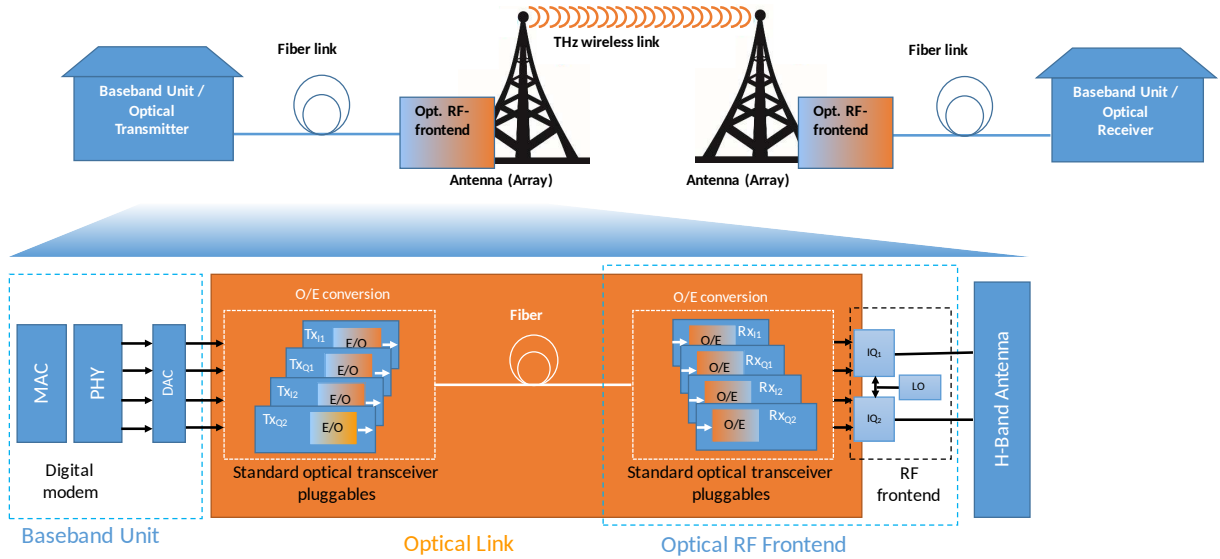}
\vspace{-3cm}
\caption{General architecture of the photonic radio.}
\label{fig:System}
\end{figure}

The generic {concept} of a photonic radio is presented in Fig.~\ref{fig:System} for a backhaul application scenario. {This system concept can be directly  applied to wireless access~scenarios.}

The digital baseband unit and the optical RF {frontend} are connected with an optical-link using coherent optical transceiver pluggable at the E/O interface. The coherent optical-fiber transmission exploits  two polarizations per optical wavelength; each polarization transmits a {QAM} signal. At the baseband, two in-phase and quadrature (IQ) signal components are mapped to the RF frontend. In comparison to RoF, the modulated RF carrier is not modulated on the optical carrier, which would require ultra-broadband optical modulators and photo-diodes to be developed. In contrast, {photonic radios}  use the baseband infrastructure of native fiber-optical systems and maps the baseband data signals between the RF frontend and the optical transponder. Apart from performance aspects, the use of optical components-off-the-shelf with minor modifications and the integration of analog interfaces with THz frontend circuit technologies are some of {their} most appealing~features.

\subsection{Baseband Interface Architectures}

\begin{figure}
\centering\includegraphics[width=1\linewidth,trim=0 0 0 0,clip=false]{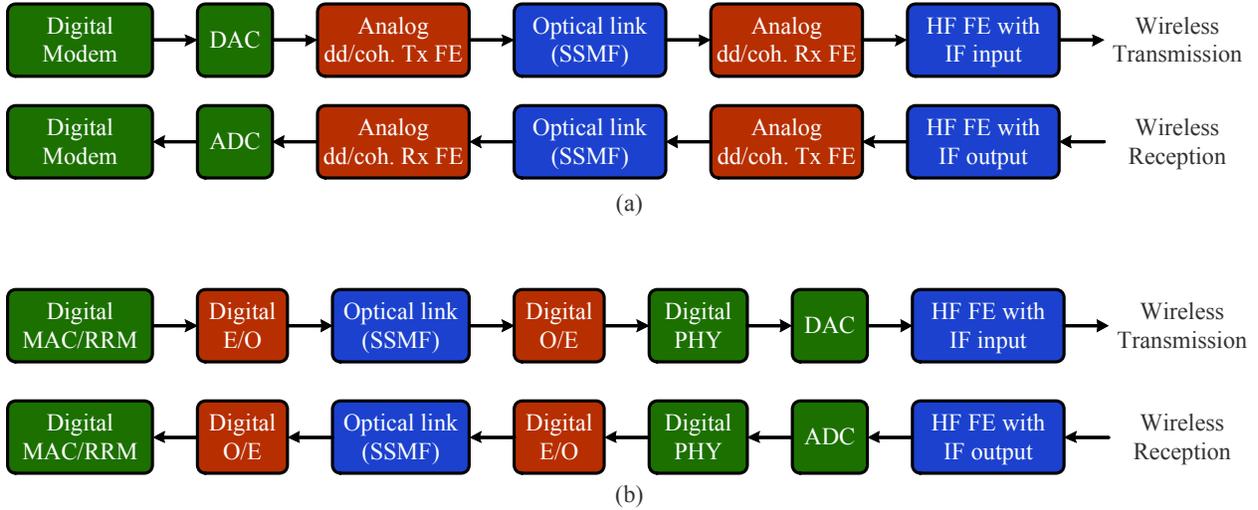}
%\vspace{-1cm}
\caption{Transceiver architectures: a) Transparent optical-link architecture, b) Digital optical-link~architecture.}
%\hrulefill
\label{fig:TRX}
\end{figure}

It is instructive to compare two different architectural variations, {namely `transparent optical-link architecture'  and  `digital optical-link architecture'} for the implementation of the photonic radio in order to {present} their basic similarities and differences. {These variations are graphically illustrated in Fig.~\ref{fig:TRX}.} 

The transparent optical-link is characterized by the conversion between optics and wireless with no additional DSP elements. The conversion step is not visible to higher layers in the protocol stack. This concept might be in principle realized by using commercially available analog coherent optics pluggable modules~\cite{OIF}. In this case, the digital modem has to cover both optical and wireless link impairments, i.e., requiring DSP co-design at the physical (PHY) layer~\cite{7193479}. A first challenge is the shared link budget between optic and wireless, which will most likely limit the available transmission distance and/or capacity. The second challenge is the geographical distance between the digital modem and the antennas, which will increase the latency and might restrict the {adaptability} and tracking speed. The third challenge is the analog mapping between optic and wireless signals, {which will be an interesting problem to be investigated; particularly, in the case of MIMO systems.} {Moreover}, the concept of the transparent optical-link potentially allows the integration of the optical/wireless frontend into a single analog module with superior compactness and power~consumption.

The digital optical-link architecture uses pluggable optical modules with Ethernet-compliant interfaces. 
This will restrict the optical-link distance and capacity to the standardized specifications~\cite{alliance20162016}. 
Additionally, a dedicated PHY will be required close to the THz antennas, increasing the complexity and power consumption of the wireless system. The DSP co-design between optics and wireless will be limited to MAC and radio resource management functions. 
The resulting DSP architecture {has several} similarities to the DSP functional split for 5G converged networks~\cite{ITG}. The optical-link architecture is generally applicable to any Ethernet-based point-to-point and networking scenario including statistical multiplexing, {given that} the QoS can be~guaranteed.

The implementation of photonic radios is {not} limited to either of the presented system architectures, as both have advantages and disadvantages; even combinations may be possible. {Therefore,}  it is important to implement proof-of-concept demonstrators for different application scenarios  to {identify} the most suitable architecture. 

\subsection{THz Wireless Link Budget}

\begin{figure}
\centering\includegraphics[width=1.0\linewidth,trim=0 0 0 0,clip=false]{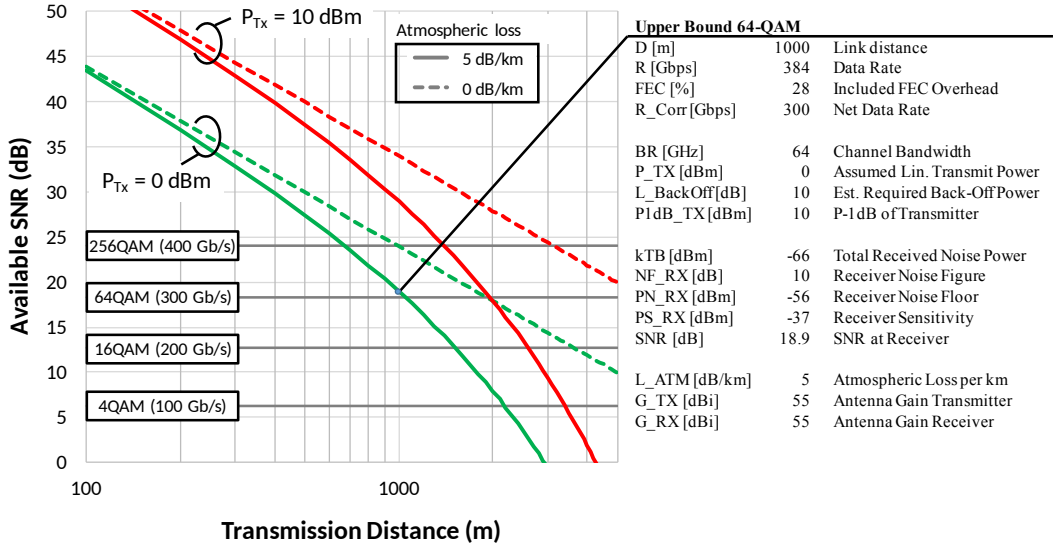}
\vspace{-2cm}
\caption{Estimated upper net data-rate bounds on THz wireless links (assuming forward error correction (FEC) threshold at BER = $2\times 10^{-2})$.}
\label{fig:distSNR}
%\hrulefill
\end{figure}

The optical wireless link-budget mostly determines the possible {applications} of photonic radios. For short-{range} optical-link of a few kilometers, the THz wireless link budget sets an upper-bound on the overall link capacity. The estimated limits on the data-rates and distances at the frequency of $300$ GHz and a gross symbol rate of $64$ Gbd are depicted in Fig.~\ref{fig:distSNR}. 

{The calculations assume a single-channel frontend having a receiver noise figure (NF) of $10$~dB over a bandwidth of at least $64$~GHz, and a linear transmit power of $0$~dBm, which are values close to those derived from latest experimental demonstrator modules and systems at 300 GHz~\cite{tessmann2017,C:Testbed}. 
Although the upper-bound calculation is based for simplicity on ideal transmitter and receivers without additional impairments like I/Q imbalances or phase noise, this is justified by assuming the use of impairments mitigating DSP, which can be effectively used to idealize optical transceivers~\cite{Pfau,morero2016}. The uncorrected amplitude and phase mismatches are in the order of $1$~dB and $1^o$, respectively~\cite{tessmann2017}. The phase imbalance was estimated from on-chip measurements of the hybrid components used in the transceiver chip.}  A highly directive antenna  with a gain of $55$ dBi is assumed at both {link ends.} This corresponds to a physical antenna aperture size having a diameter of $225$ mm and an aperture efficiency of $80 \%$. Currently available broadband III-V semiconductor frontend technologies can support a transmit power of $0$ dBm  backed off into the required linear operating regime. {Extrapolating on} the technological progress within the next {few} years, integrated frontends are in reach that will be able to transmit at a linear power level of $10$ dBm. The actually required backoff power is an implementation factor, influenced by linearization techniques and coding, which play  important roles in covering longer-range THz links. {This figure} also highlights {the importance of} accurately {modeling} the free space propagation losses caused by atmospheric attenuation. The current III-V receiver performance limits the order of the QAM to 128, {due to the tradeoff between mixer linearity and receiver NF. The upcoming optical transponder chipsets will support up to 64-QAM. This means that a current photonic radio link has the potential to support a maximum data-rate of approximately $300$ Gbps over a wireless distance of $1$ km. By utilizing two antenna polarizations, a theoretical upper limit of $600$ Gbps can be derived. Taking into account the predicted progress in coherent optical systems towards 128-QAM~\cite{roberts2017}, maximum data-rates per link of $800$ Gbps using two antenna polarizations will be feasible within the next few years}. The data-rate compatibility between the optical and THz  link of {the} photonic radio is a plausible vision, even {without considering} neglecting the additional potential of full LOS MIMO~architectures.

\section{Design Principles And Technology Enablers}

%The previous section has shown that the development and implementation of photonic radio architectures, especially the transparent optical-link architecture, entails several interdependent challenges. 
{The design of THz wireless networks benefits from the principle of co-design of new signals, codes and protocols along with  THz channel/propagation models and pencil-beam antenna arrays.}
%THz fontends and interfaces capable to support Tbps data-rates need to be developed and integrated together. For THz systems{,} advanced massive MIMO processing methods for pencil-beam antennas will gain more {attention}, and hybrid multi-user/device {MAC} protocols need to be designed, {that take into consideration the high antenna directionality}. Finally, it is necessary to derive novel approaches in optimization theory and machine learning for achieving optimal spectrum~utilization.
{Next, some of the enabling technologies are presented in more detail. The channel characteristics for frequencies below $1$ THz and the advances in implementing frontend architectures for frequencies above 275 GHz are discussed, and possible THz beamforming (BF) architectures are reviewed.}

\subsection{THz Channel}

{THz channel models are}  important not only as part of the first step for developing {the} application scenarios{,} but also for dimensioning phased array configurations and transceiver architectures{, during the} implementation phase. 
For example, {for} wideband channels from $275-325$ GHz, it is of interest to pre-equalize frequency-dependent path-loss in the vicinity of spectral atmospheric absorption maxima at the THz frontend. 
This becomes practically relevant for link-distances of several hundreds of meters, especially when {highly spectral efficient modulations} are employed. {Note that this} is  a typical scenario for small-cell backhaul links. 
One of the most intriguing new aspects of photonic radio design is the development of combined optical-wireless channel models for supporting the E2E error-correction~idea. 

\begin{figure}
\centering\includegraphics[width=1\linewidth,trim=0 0 0 0,clip=false]{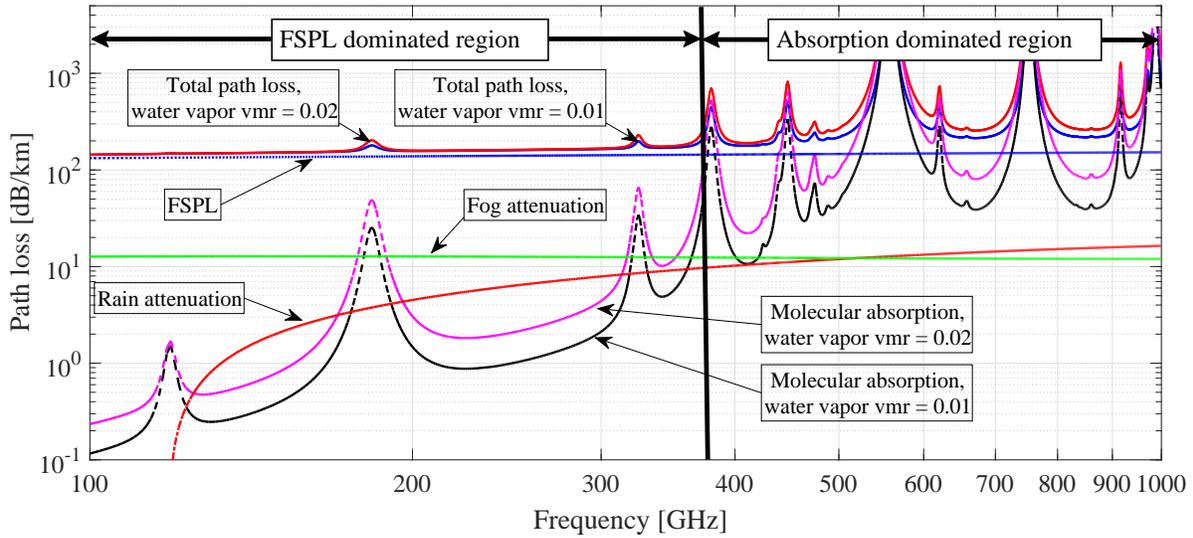}
\caption{Individual loss components of the lower THz band LOS channel, as well as the expected total losses under harsh weather conditions. {The absorption losses and the FSPL were calculated with the HITRAN-based line-by-line model and Friis equation, respectively.}}
\label{fig:Channel}
\end{figure}

Apart from the combined channel modeling problem, the THz link suffers from several path-loss mechanisms, {including} the free space path-loss (FSPL), due to  {signal spreading,} the effective antenna aperture, and the molecular absorption. The latter is a distinguishing feature of the millimeter and submillimeter bands. {The main difference between the mmW  and the THz band is the progressively increasing molecular absorption loss. At longer link distances, below $300$ GHz, links are dominantly attenuated by the FSPL, whereas in the THz regime, the molecular absorption becomes more important, due to its exponentially increasing impact as a function of distance. At short distances (tens of meters), the FSPL remains the dominant loss.} The existence of the molecular absorption loss highly depends on the level of the water vapor in the atmosphere, since, in this band, water vapor efficiently absorbs energy. This indicates that there are global variations in the level of absorption. {Therefore}, for~standardization, it is important to take into account the regional propagation properties.

Fig.~\ref{fig:Channel} compares the possible LOS losses per kilometer and {their} total contributions. As the distance between the {transceiver nodes} increases, the FSPL dominates the total loss below $370$ GHz, whereas above $370$ GHz, the molecular absorption loss dominates.  Even below $370$ GHz, {we observe} three absorption lines, {which} depends on the distance and humidity level. Above 370 GHz, molecular absorption has clearly larger effect, but, depending on the scenario and frequencies considered, below 370 GHz molecular absorption may also have to be taken into account. 
Below the $1$ THz, there are several transmission windows exhibiting minimum molecular absorption that can be utilized in long-distance~links. 

The absorption losses were calculated based on spectroscopic databases (HITRAN) in conjunction with the Beer-Lambert law. {They} were calculated for {water vapor} volume mixing ratios $0.01$ and $0.02$, at the Earth's surface level, representing roughly the mean humidity in Europe and the equatorial regions in June 2016, respectively. Because of the exponential nature of the molecular absorption and the volume mixing ratios of the molecules, which is used to weight the total path-loss of the mixture, doubling the amount of water vapor roughly doubles the loss on a dB scale. ITU-R Recommendations P.838-3 and P.840-6 were used for evaluation of the attenuation caused by rain and fog, respectively. In Fig.~\ref{fig:Channel}, {the curves} correspond to a heavy rain of $50$ mm/hr and a dense fog of $0.5$ $\mathrm{g/m}^3$ liquid water content in air. These conditions cause an additional approximately $10$ dB loss. However, under normal conditions, the total path-loss can be expected to be $10-20$ dB smaller than the heavy rain and deep fog~cases.

\subsection{THz Transceiver Frontend}

The design of transceivers for broadband transmission in the THz  regime brings several challenges, which are of technological nature, but also related to the art of circuit implementation. {Before discussing the transceiver~challenges,}  {we revisit the different available semiconductor transistor options and materials} for the integration of `all-electronic' frontends above $200$ GHz, {in order to present} some of the basic trends.

For more than a decade, III-V compound semiconductor based high-electron-mobility transistor (HEMT) and heterojunction-bipolar-transistor (HBT) technologies have dominated most of the frontend integrated {circuits.} The major reasons are the exceptional power and low-noise capabilities of the devices in comparison to other options. {Recently}, there has also been  significant progress reported in silicon complementary metal-oxide-semiconductor (CMOS) and silicon germanium (SiGe) HBTs. While SiGe HBTs have achieved progress in power capabilities and bandwidth, first {integrated front end solutions for wireless communications at $240$ GHz} were reported in 2016, the CMOS high frequency transistor based solutions still lack in bandwidth. Multichannel transceiver architectures dividing the $275-325$ GHz spectrum into different sub-bands may be a variable approach to {countermeasure} the CMOS bandwidth limitations. On the downside, this {approach} increases the frontend complexity. When additional beamsteering capabilities {are required,} broadband amplifiers will be necessary to compensate power divider~losses. 

Up to now, SiGe and {CMOS-based} frontends have focused on THz short-range {communications,} employing integrated on-chip antennas. {These on-chip antennas could be also used for feeding high-gain reflector type or lens-type antennas,  which are essential to address THz backhaul requirements.} Backhaul {requirements} have been almost-exclusively addressed by integrated frontends using III-V HEMT or HBT transistors as of today. {Only} recently, {devices operating} at $300$ GHz, exhibiting bandwidths of more than $50$ GHz, were realized~\cite{C:Testbed}.

For all transistor technology options, direct conversion {transceiver} architectures are currently the most widespread, due to their low integration complexity. However, when targeting the use of higher-order modulations, those architectures suffer from imperfections that need to be {mitigated by baseband DSP~\cite{roberts2017,morero2016, Pfau}. Most critical is the correction of IQ imbalances, amplifier nonlinearities and local oscillator (LO) phase noise~(PHN). }

{We highlight that the carrier phase recovery in the presence of PHN by feedforward algorithms become more robust for higher symbol rates \cite{Pfau}. While the LO PHN degrades with frequency, for example in~\cite{C:Testbed}, LO PHN at $300$ GHz of $95$ dBc/Hz at $100$ kHz was evaluated from measurements, the thermal phase noise floor becomes the limiting factor for the high symbol rates, as concluded in~[5]. 
The thermal PHN floor is mainly dominated by the carrier power,  which explains why IQ mixers that require high local oscillator power levels are beneficial at $300$ GHz. }

The error-correction schemes and frontend requirements in photonic radio systems are {a} subject {for} research. 
{In recent years, there has been great progress reported in developing mitigation schemes for carrier PHN, nonlinear dispersion, and IQ imbalance  for coherent fiber optical systems enabling $64$-QAM in future systems~\cite{morero2016}}. 
The baseband interface plays an important role, since it needs to handle symbol rates of up to $64$ Gbd, support at least $64$-QAM, map the symbols to the optical-fiber electronics and {take} care of pre-equalization. 
{This is in-line with the advances of high-speed DACs, ADCs and specialized DSP ASICs~\cite{laperle2014}, reaching sampling rates of $90$ GS/s and analog input bandwidths of more than $20$ GHz with $8$~bit nominal resolution and up to $6b$~effective number of~bits}. 
%Most of the first generation THz frontends did not employ specially designed baseband interfaces, since they were only tested by lab equipment in conjunction with offline post-processing of captured~data. 

{Finally,} the optical interface poses new research {challenges} on packaging solutions. While the monolithic microwave integrated circuits can be easily tested in waveguide packages, integration with optical transceivers in one single-package (e.g., the Quad Small Form-Factor Pluggable Plus) brings {several} technological challenges, but also a great {opportunity} to reduce overall system costs and form suitable factors for {commercialization}~{\cite{roberts2017}}.

\subsection{Pencil Beamforming and Resource Management}

BF architectures, which {are} especially attractive for short-range links to address wireless access scenarios, can be employed for compensating the FSPL.
Due to the small wavelengths and the related small-sized antenna elements, a large number of antenna elements could be installed to form powerful BF systems. {Such BF} architectures for THz communications have been outlined in~\cite{A:THzCommunicationsAnArray_of_Subarrays_Solution}. {However, fully surface emitting 2D-array architectures in backhaul links will remain costly for a long time.} For example, approximately $200$ active elements will be required to replace a $50$ dBi reflector antenna at the transmitter. While this integration density might be within the roadmap of CMOS technologies, the power consumption of such arrays will also be increased by a factor of $200$. {Additionally}, {advanced} 3D integration methods will be required to realize hybrid 2D array antenna architectures and routing out multiple channels at data-rates of up to $400$ Gbps from an integrated frontend chip. For this reason, the reduction of the required number of active antenna elements will be one of the most important technology enablers. The antenna element pitch of the array poses limits on the architecture of the {transceivers} and their integration at chip, module and system level. Many of those basic considerations and trade-offs have been intensively investigated in upcoming 5G scenarios at $28$ GHz for LOS and non-LOS~\cite{7342886}.  

\begin{figure}
\centering\includegraphics[width=0.74\linewidth,trim=0 0 0 0,clip=false]{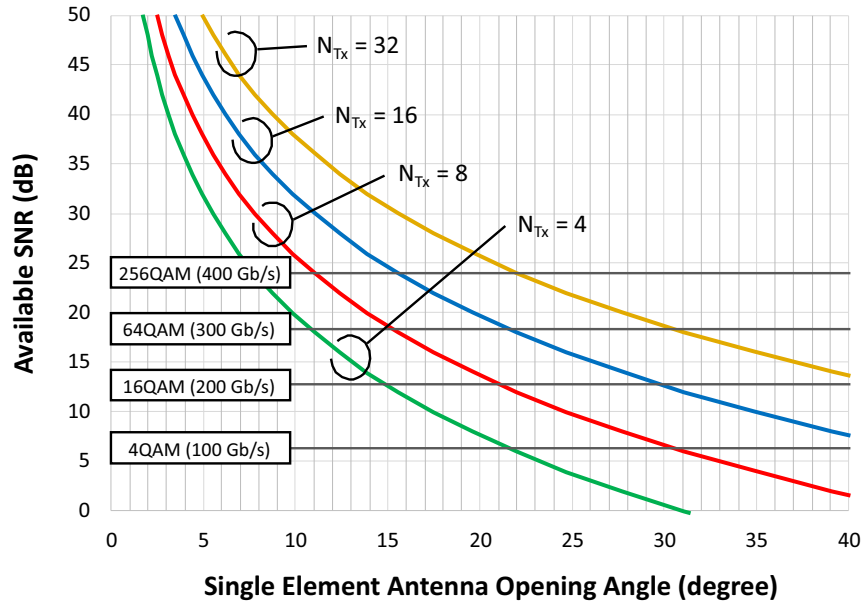}
\caption{Estimated data-rates for small-sized THz sub-arrays.}
\label{fig:dataRate}
\end{figure}

In Fig.~\ref{fig:dataRate}, the maximum antenna opening angle for small-sized sub-arrays composed of $N_{TX}$ antenna elements is~{depicted}. The antenna opening angle corresponds to $2\times$ the maximum beam-scanning angle from broadside direction.  The link distance is fixed to $10$ m and the linear transmit power to $0$ dBm per antenna~element. Apart from the antenna gain and spatial power combining benefit all other assumptions are equivalent to the backhaul scenario, {provided in Fig.~\ref{fig:distSNR}}.

The calculations in Fig.~\ref{fig:dataRate} assume that the array opening angle corresponds to the $3$dB-beamwidth of a single-antenna element, {which amounts to an upper-bound of the array factor}. This figure predicts that data-rates up to $200$ Gbps {can} be possible with a {maximum} antenna opening angle of $15^o$, when using $4$  elements. 
To increase the usefulness, in practical applications{, $N_{Tx}$ needs} to be increased to at least $8$ or $16$, which allows a data-rate increase or {improves the} trade-off {between} data-rate {and} opening angle. A first experimental $4$-channel platform was developed in~\cite{C:Testbed}, for verifying this trade-off  at $300$ GHz. {For the envisioned scenarios of limited number of antenna elements between $4$ and $16$  and scanning angles of less then $\pm 25^{o}$, beamsteering can be achieved by either time-shifting in the digital baseband, phase-shifting in the LO path or in combination, at relative bandwidths of $20\%$.} 

%\subsection{PHY \& MAC}

{By exploiting the capabilities of the THz channel particularities, transceivers and BF, new PHY challenges and opportunities are revealed, which include the design of modulation and coding (MC), synchronization, and equalization techniques. 
Since the bandwidth drastically changes even with small variations in the transmission distance, the development of distance-aware MC schemes that take into account the distance-dependent bandwidth is required. Hence, distance-adaptive MC should be developed that employ either (i) the entire transmission bandwidth (for short-distance communications), (ii) the central part of the transmission window (to serve close and far nodes), or (iii) the sides of the transmission window (for short-distance links).}

{PBF requires accurate channel state information  to intelligently adapt the antenna array. However, as the number of antennas increases, channel estimation becomes unfeasible. Additionally, if individual transceivers are equipped with single-antenna, cooperative BF can be used; however, this comes with the challenge of tight synchronization and coordination among neighbor nodes.}

{In cooperative communications, since each node has independent LOs, significant frequency offsets across devices can be generated.  Hence, the development of accurate synchronization mechanisms is required for ultra-broadband THz communications. In this sense, an important research challenge is to invent novel schemes for time synchronization.
Another  challenge is to utilize novel mechanisms for accurate frequency synchronization to eliminate the frequency offset among different nodes. 
Finally, in the case of  PBF, where the beam-widths are very narrow, accurate pointing and automatic beam steering schemes are necessary. In other words, low-complexity space synchronization mechanisms need to be designed that  minimize the required overhead.}

{Achieving reliable Tbps connectivity in beyond 5G THz  systems largely} depends on a specialized MAC design for unique THz band properties and {application} scenarios. When a large number of devices simultaneously communicate,  high data-rate transmission,  zero-latency requirements,  pencil-beam antennas,  super-wide available bandwidth and  {all other} peculiarities of the THz band result in new THz channel and interference models, which render the classical MAC protocols  inadequate and not directly applicable. The {use of PBF, in order to expand the range of the THz link,}  requires new methods for device discovery{/tracking, and} resources management under the constraint of the absence of an omnidirectional broadcast channel. These methods should be able to cope with distance-dependent bandwidth and new interference models, by capitalizing on extreme densification and hybrid/flexible deterministic/random access principles. {While mmW systems, such as IEEE 802.11ad and Wireless HD may use quasi-omnidirectional antenna patterns for discovery and signaling messages, THz band protocols will not have this luxury; thus new MAC designs are needed.} Finally, {the efficient exploitation of} the enormous  bandwidth in multi-user/device networks, {requires} a strategic spectrum utilization design, which simultaneously supports multiple ultra-high-speed links.

\section{Conclusions}

{We} presented {the} concept for blending wired-optical and wireless-THz links, its requirements, and possible applications. The development of the THz  system and baseband interface is in its early stage; hence, it is difficult to {specify} precisely the final system {architecture and features.} However, it is  possible to identify  the enablers that will boost {its} utilization. {Specifically, in this article, we highlighted these enablers, namely} channel modeling, {the development of novel PHY and MAC protocols, the design of transceiver RF frontend and baseband DSP algorithms to mitigate the hardware constraints, and the implementation of novel PBF schemes.} Finally, the co-design principle and important research directions {were~discussed}.

\section*{Acknowledgment}

The authors would like to thank all the colleagues of the project TERRANOVA for their support. This work has received funding from the European Commission's Horizon 2020 research and innovation programme under grant agreement No~761794.

\balance
\bibliographystyle{IEEEtran}

\begin{IEEEbiography}{Alexandros--Apostolos A. Boulogeorgos} (S'11-M'16)  received the Electrical and Computer Engineering diploma and PhD degrees both from the Aristotle University of Thessaloniki in 2012 and 2016, respectively. From March to August 2016, he was a postdoctoral researcher in the Centre for Research and Technology Hellas, while, from October 2016, he works as a postdoctoral researcher in the University of Piraeus. His research interests include wireless and THz communication theory, as well as optical wireless communications.      
\end{IEEEbiography}

\begin{IEEEbiography}{Angeliki Alexiou} received the Diploma and PhD in Electrical and Computer Engineering from the National Technical University of Athens and the Imperial College, University of London in 1994 and 2000, respectively. Since 2009, she is faculty member at the Department of Digital Systems, University of Piraeus. Prior to this, she was with Bell Laboratories, Lucent Technologies in UK (1999-2009). Her research interests include high frequencies technologies, and resource management for Ultra Dense wireless networks.
\end{IEEEbiography}

\begin{IEEEbiography}{Thomas Merkle} received his M.Sc. and PhD degree in Germany from the University of Stuttgart, and the University of Ilmenau, in 1999 respectively 2006. From 2005 to 2010, he worked as a Post-Doctoral Fellow at the CSIRO ICT Centre, Sydney, Australia. From 2010 to 2013, he was a Senior Research Engineer at the Sony STC in Germany. Currently, he is with the Fraunhofer IAF, focusing on THz wireless communications and related device technologies.
\end{IEEEbiography}

\begin{IEEEbiography}{Colja Schubert} received his PhD in physics from Technische Universität Berlin, Germany, in 2004.
Since 2000, he is a member of the scientific staff at Heinrich-Hertz-Institute, Berlin. His research interests include high-speed transmission systems and all-optical signal processing. He is head of the Submarine and Core Systems group and deputy department head of the Photonic Networks \& Systems Department.
Dr. Schubert is a member of the German Physical Society.
\end{IEEEbiography}

\begin{IEEEbiography}{Robert Elschner} received the Dipl.-Ing. and Dr.-Ing. degrees in electrical engineering from Technische Universit\"at Berlin, Germany in 2006 and 2011, respectively. In 2005, he was with T\'el\'ecom ParisTech. 
Since 2010, he is a member of the scientific staff and project manager at the Fraunhofer Heinrich Hertz Institute, Berlin, Germany, working in the field of digital coherent optical transmission technology.
Dr. Elschner is currently serving in the Technical Program Committee of the Optical Fiber Communication Conference.
\end{IEEEbiography}

\begin{IEEEbiography}{Alexandros Katsiotis} received the B.Sc. degree in physics, the M.Sc. degree in telecommunications, and the Ph.D. degree in coding theory from the University of Athens, Greece, in 2003, 2005, and 2012, respectively. From 2012 to 2016 he was a Research Associate with the Department of Informatics and Telecommunications, University of Athens. Since 2016 he is with Intracom S.A. Telecom Solutions. His research interests are in the area of wireless communications, coding theory and applications. 
\end{IEEEbiography}

\begin{IEEEbiography}{Panagiotis Stavrianos} received the BSc degree from Electronics Engineering department of Piraeus University of Applied Sciences, Greece. Since April 2016, he is a Junior Wireless Systems Engineer at the R\&D division of INTRACOM S.A. TELECOM SOLUTIONS, Greece working on digital signal processing algorithms for microwave and mm-wave broadband wireless systems. His research interests include channel characterization, physical layer techniques for wireless access and transmission systems, adaptive equalization, symbol \& carrier synchronization, interference cancellation, beamforming and dirty RF compensation techniques.
\end{IEEEbiography}

\begin{IEEEbiography}{Dimitrios S. Kritharidis} received his Diploma in Electrical Engineering from National Technical University of Athens, Greece in 1988 and his MSc in Microprocessor Engineering and Digital Electronics from UMIST, U.K. in 1989. After a year in Siemens Medical Division, in 1991 he joined Intracom Telecom as an ASIC Designer and from 1993-2009 he led the IC Design Group. Currently he is leading the research activities of the Wireless \& Network Systems BU of the company.
\end{IEEEbiography}

\begin{IEEEbiography}{Panteleimon-Konstantinos Chartsias} received his Diploma in Electrical and Computer Engineering from National Technical University of Athens (NTUA), Greece in 2015. He is a member of Internet Systematics Lab (ISLAB) of N.C.S.R Demokritos and he is currently working at INTRACOM S.A. TELECOM SOLUTIONS as an R\&D engineer where he is mainly involved in CHARISMA and SPEED-5G, two 5G PPP H2020 projects. His research interests include Software Defined Networks, broadband networks, network security and smart grid. 
\end{IEEEbiography}

\begin{IEEEbiography}{Joonas Kokkoniemi} received Dr.Sc. degree in communications engineering from the University of Oulu, Oulu, Finland, in 2017. He is a postdoctoral researcher with the Centre for Wireless Communications, University of Oulu. From September to November 2013, he was a Visiting Researcher with the Tokyo University of Agriculture and Technology, Tokyo, Japan. From March to October 2017, he was a Visiting Postdoctoral Researcher with the University at Buffalo, NY, USA. His research interests include THz band channel modeling.

\end{IEEEbiography}

\begin{IEEEbiography}{Markku Juntti} (S'93-M'98-SM'04) received doctorate in 1997. He was a Visiting Scholar at Rice University, Houston, Texas in 1994--95, and a Senior Specialist with Nokia Networks in 1999--2000. He has been a professor of communications engineering since 2000 at University of Oulu, Centre for Wireless Communications (CWC), where he serves as Head of CWC -- Radio Technologies (RT) Research Unit. His research interests include signal processing for wireless networks, communication and information theory.
\end{IEEEbiography}

\begin{IEEEbiography}{Janne Lehtom\"aki} got his doctorate from the University of Oulu, Finland, in 2005. Currently, he is adjunct professor (docent) at the University of Oulu, Centre for Wireless Communications. He spent the fall 2013 semester at the Georgia Tech, Atlanta, USA, as a visiting scholar. Currently, he is focusing on terahertz band wireless communication. He co-authored the paper receiving the Best Paper Award in IEEE WCNC 2012. He is associate editor of Physical Communication.
\end{IEEEbiography}

\begin{IEEEbiography}{Ant\'onio Teixeira} got his PhD from university of Aveiro in 1999. From 1999 is a Professor in Universidade de Aveiro. Since 2014, he is the Dean of the University of Aveiro Doctoral School. In 2014 he co-founded PICadvanced. He has served the ECOC TPC from 2008-15 and he has served the access subcommittee in OFC from 2011-14. He is a Senior Member of OSA and a member of IEEE and IEEE standards association
\end{IEEEbiography}

\begin{IEEEbiography}{Francisco Rodrigues}
received Masters in Electronics and Telecommunicatios from Universidade de Aveiro, Portugal in 2014. In September 2014 starts the PhD program in Electrical Engineering and works as Optics Engineers in PT Inovac\~ao, S.A. In September 2015 joins PICadvanced as Optics Engineer where he is currently acting as CEO.
\end{IEEEbiography}

\end{document}